\documentclass[prb,showpacs,floatfix,twocolumn]{revtex4-1}
\usepackage{graphicx}
\usepackage{amssymb}
\usepackage{dcolumn}
\usepackage{bm}

\usepackage{color}

\usepackage{subfigure}

\begin{document}

\title{Seismic cycles, size of the largest events, and the avalanche size 
distribution in a model of seismicity
}

\author{L. E. Arag\'on and E. A. Jagla}

\affiliation{Centro At\'omico Bariloche and Instituto Balseiro, Comisi\'on Nacional de Energ\'{\i}a At\'omica, 
(8400) Bariloche, Argentina}

\author{A. Rosso}

\affiliation{CNRS-Laboratoire de Physique Th\'eorique et Mod\`eles Statistiques, 
Universit\'e Paris-Sud, 91405 Orsay, France
}

\begin{abstract}
We address several questions on the behavior of a numerical model recently introduced to study seismic phenomena, that includes relaxation in the plates 
as a key ingredient. We make an analysis of the scaling of the largest  events with system size, and show that when parameters are appropriately interpreted, the typical 
size of the largest events scale as the system size, without the necessity to tune any parameter. Secondly, we show that the temporal activity
in the model is inherently non-stationary, and obtain from here justification
and support for the concept of a ``seismic cycle" in the temporal evolution of seismic activity.
Finally, we ask for the reasons that make the model 
display a realistic value of the decaying exponent $b$ in the Gutenberg-Richter law for the 
avalanche size distribution. We explain why relaxation induces a systematic
increase of $b$ from its value $b\simeq 0.4$ observed in the absence of
relaxation. However, we have not been able to justify the actual robustness of the model
in displaying a consistent $b$ value around the experimentally observed value $b\simeq 1$.
\end{abstract}

\maketitle

\section{Introduction}

Modeling of seismic phenomena as a statistical mechanic process has a long history that goes
back almost to the very beginning of plate tectonics. There are a lot of variables 
affecting seismic activity, and many of them cannot be taken into account in a simplified theoretical description.
However, if a model gives a sequence of events strongly resembling
real ones (in its temporal, spatial, and magnitude distribution), it is tempting to consider that
the model captures 
those key ingredients playing an important role in the production of seismic events.
This is what happens with the model proposed by Burridge and Knopoff \cite{bk,langer}.
Constructed in the form of a large collection of repetitive and simple elements (rigid blocks interacting through 
elastic springs and sliding on a rigid surface), it displays non-trivial features, most remarkably
the existence of avalanches (earthquakes) with a broad distribution of sizes. 
Based on the work of Burridge and Knopoff, in 1991 Olami, Feder and Christensen\cite{ofc} (OFC) introduced
a model that has become one of the paradigms of simulation of seismic activity. The model is presented
in the form of cellular automata, and the dynamics is defined as a list of rules. A broad distribution of event sizes
is obtained with the OFC model.
Despite the attention it received, the OFC model displays a series of unrealistic features when
trying to model seismic phenomena.
For instance, the prominent spatial and temporal correlation
of events that is observed in the field, mostly the existence of aftershocks, is not reproduced\cite{ofc_aft}. Also,
obtaining a realistic value of the $b$ exponent in the Gutenberg-Richter (GR) law requires the tuning of 
an internal parameter of the model.

In the last years, we have been working in introducing modifications to this kind of models in such a way that more
realistic sequences of events are obtained. The two crucial modificationshav we have proposed 
are\cite{jagla_kolton,jagla_ofc} (see a more detailed account in the next section)
the consideration of variable thresholds (instead of the constant value considered by OFC), and the incorporation
of a mechanism of internal (also called structural) relaxation,  quantified by a parameter $R$. With these two ingredients we have obtained a model in which:
1) the spatial and temporal correlation of events is comparable to real ones, in particular, aftershock sequences obeying
the Omori law\cite{omori} are obtained, 2) the avalanche size distribution has a GR form, with an exponent $b$ very close to actually observed 
values (around 1), 
3) the friction properties derived from the model reproduce non-trivial results such as velocity weakening
or the stress peak in experiments of slip-stop-slip.\cite{vw} It is worth mentioning that as long as 
$R$ is larger than some threshold value, all these results are obtained irrespective of the precise value of $R$.

The number of realistic features that have been obtained with this model, encouraged us to go further and investigate
in more depth its properties to better understand how they originate in the existence of the relaxation process. This is the aim
of the present work.
To make the presentation self contained, in the next section the model is explained with all necessary detail, in the
case without relaxation $R=0$, and in the presence of relaxation $R\ne 0$. Also, a limiting case of infinite relaxation
$R\to\infty$ is introduced, that will be interesting to consider along the paper. 

In section III we review some results obtained in the case $R=0$. 
It is shown that this case is equivalent to the problem of depinning of an elastic interface in a random medium. We give also a comparison of the results obtained at ``constant driving velocity"
with those at ``constant force", showing that there is a clear mapping between the two.

In Section IV we show that when $R\ne 0$ we cannot make an equivalence between simulations at constant velocity and 
constant force. In fact, in constant velocity simulations at $R\ne 0$, there are systematic fluctuations of the stress on 
the system that make the problem non-stationary, and
this generates a scenario in which 
a description in terms of a ``seismic cycle" naturally emerges.

In Section V we discuss the scaling of the linear size of the largest events $L_{\max}$ with the control parameter $\alpha$.
We show that with an appropriate interpretation of $\alpha$ as inversely proportional to the thickness of the sliding slabs, 
there are events that are comparable in size with the size of the system. This is a consequence of the existence of relaxation,
and does not occur in the $R=0$ case.

Section VI is devoted to the following issue: the model with $R=0$ gives a consistent (although unrealistic) 
value for the $b$ exponent of about $b\simeq 0.4$. The inclusion of relaxation drives this exponent to
values closer to the realistic value $b\simeq 1$, without tuning of any parameters. What is the reason of this behavior?
As we will see, we are only in the position to provide a partial answer to this question.

Finally, in Section VII, we make a brief summary and present our conclusions.

\section{Models}
\subsection{OFC model}
\begin{figure}

	\subfigure{
	\includegraphics[width=0.225\textwidth]{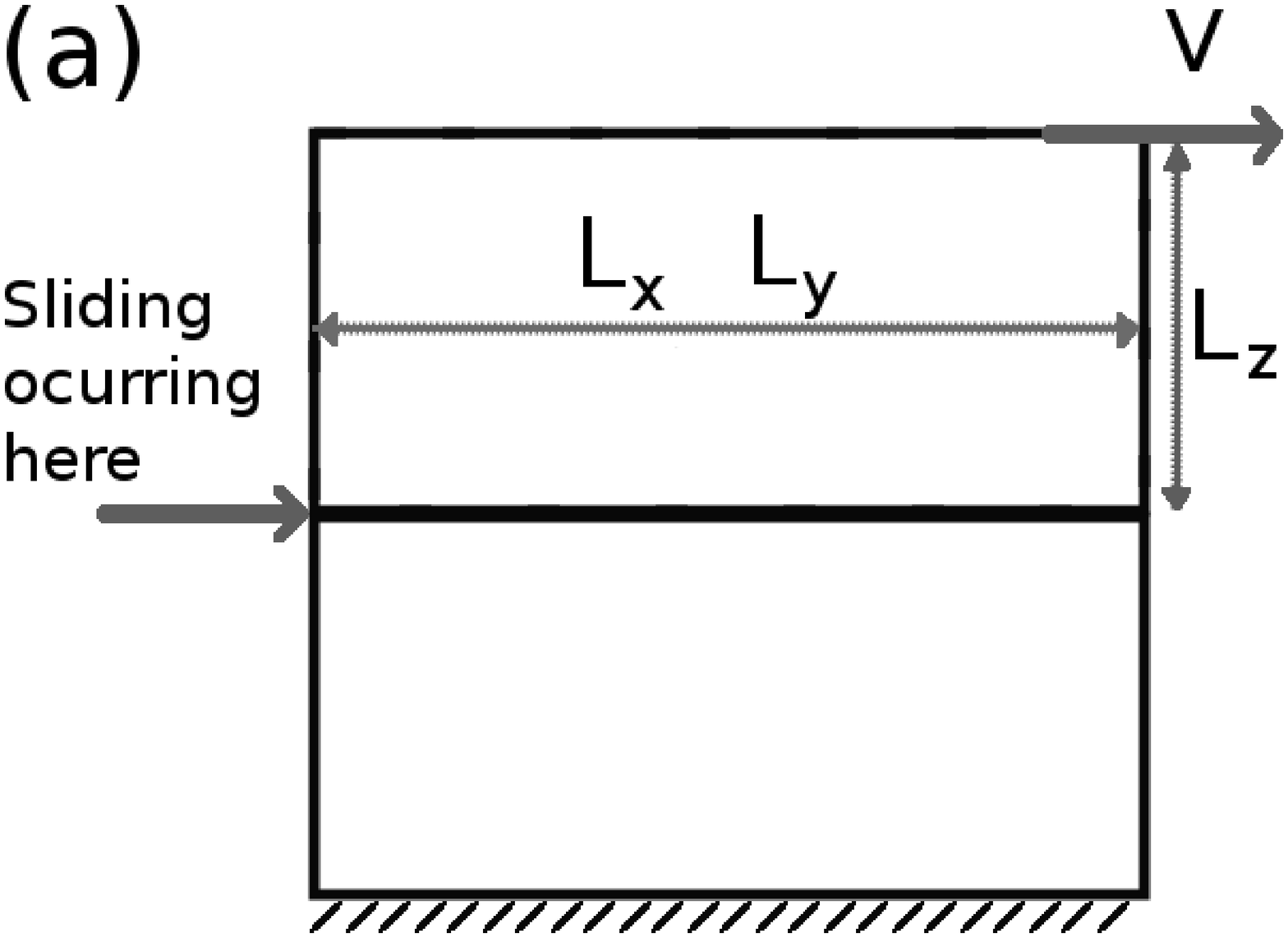}
	}
	\subfigure{
	\includegraphics[width=0.225\textwidth]{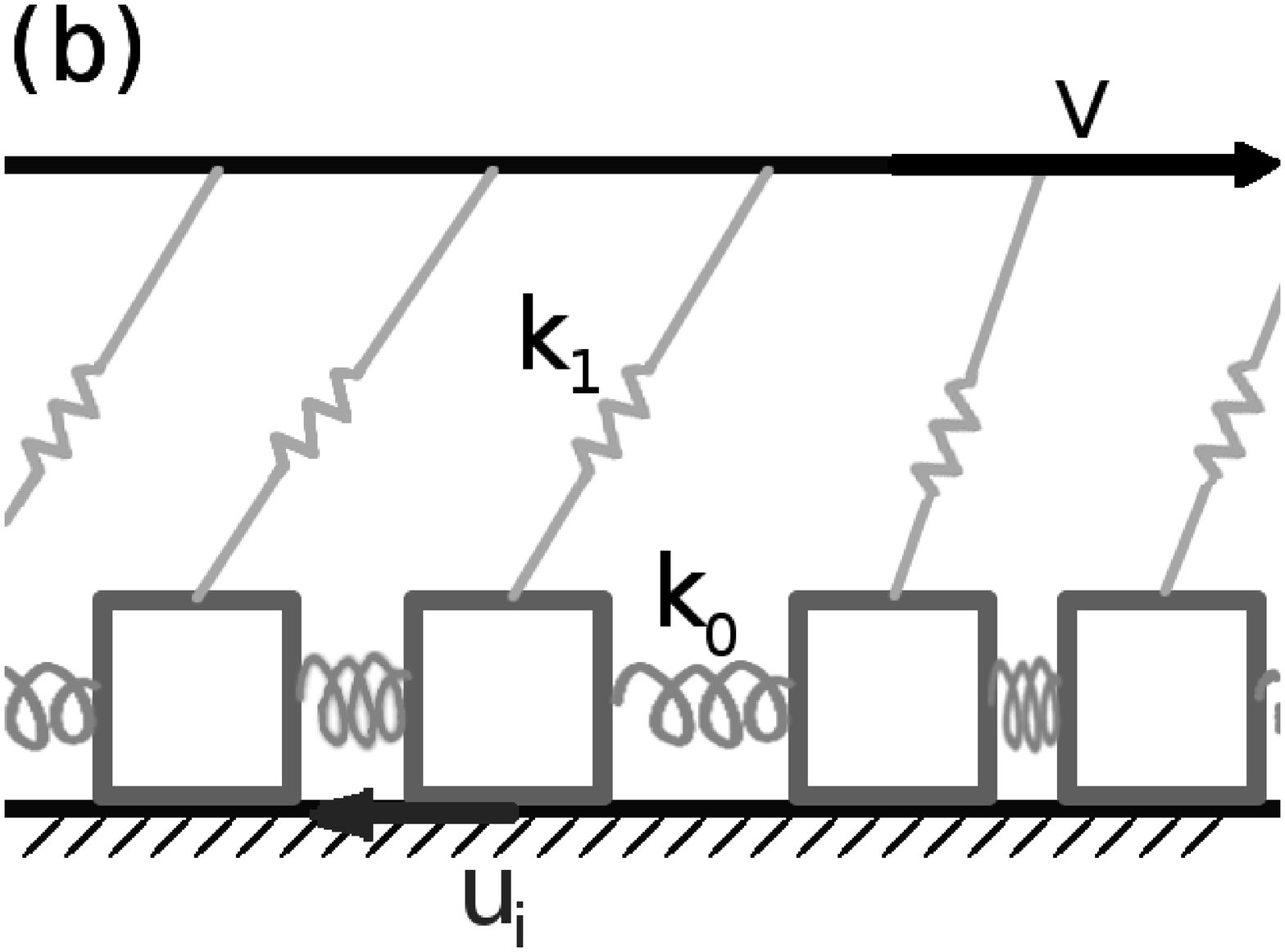}
	}	
	\subfigure{
	\includegraphics[width=0.225\textwidth]{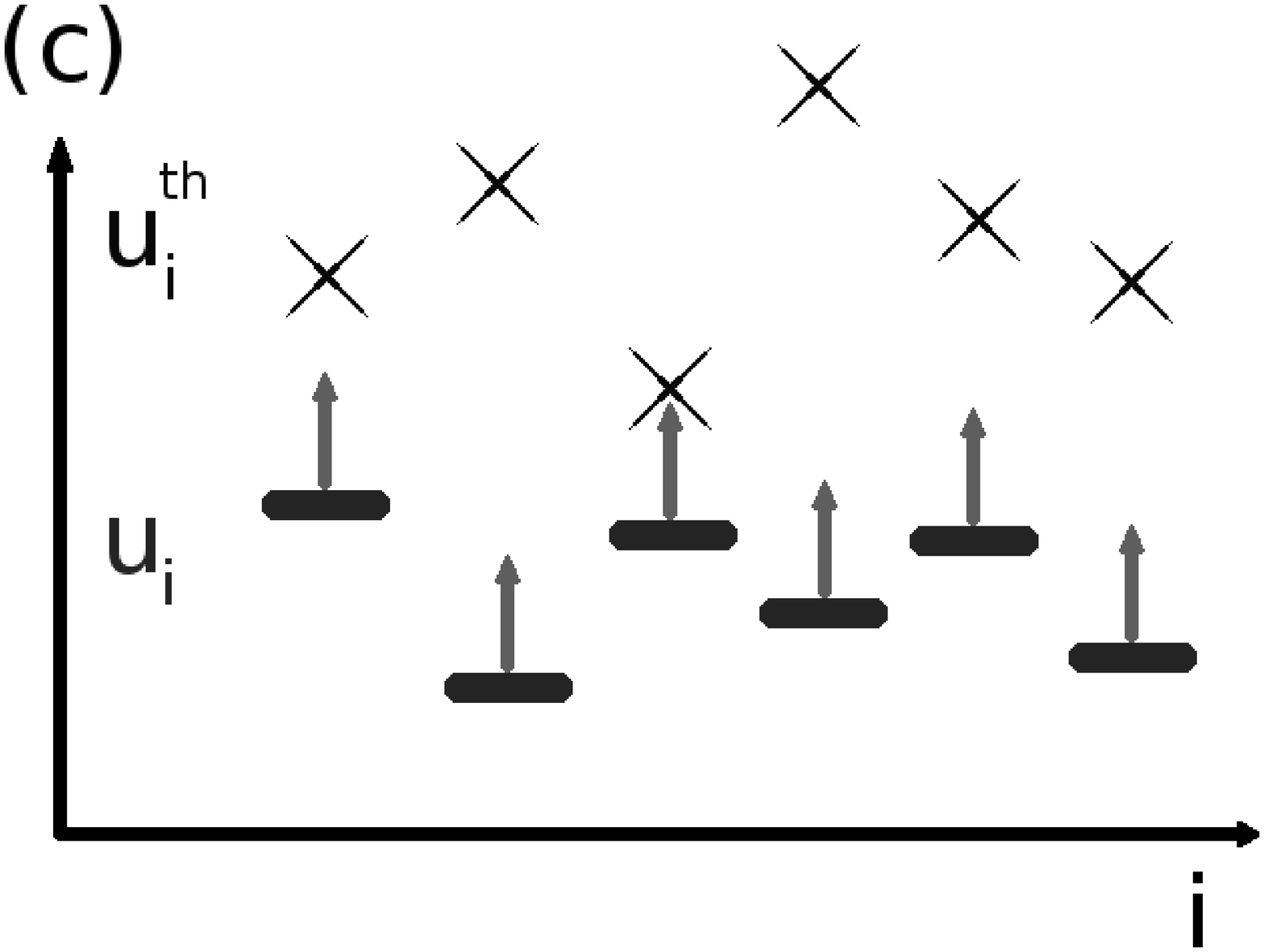}
	}
	\subfigure{
	\includegraphics[width=0.225\textwidth]{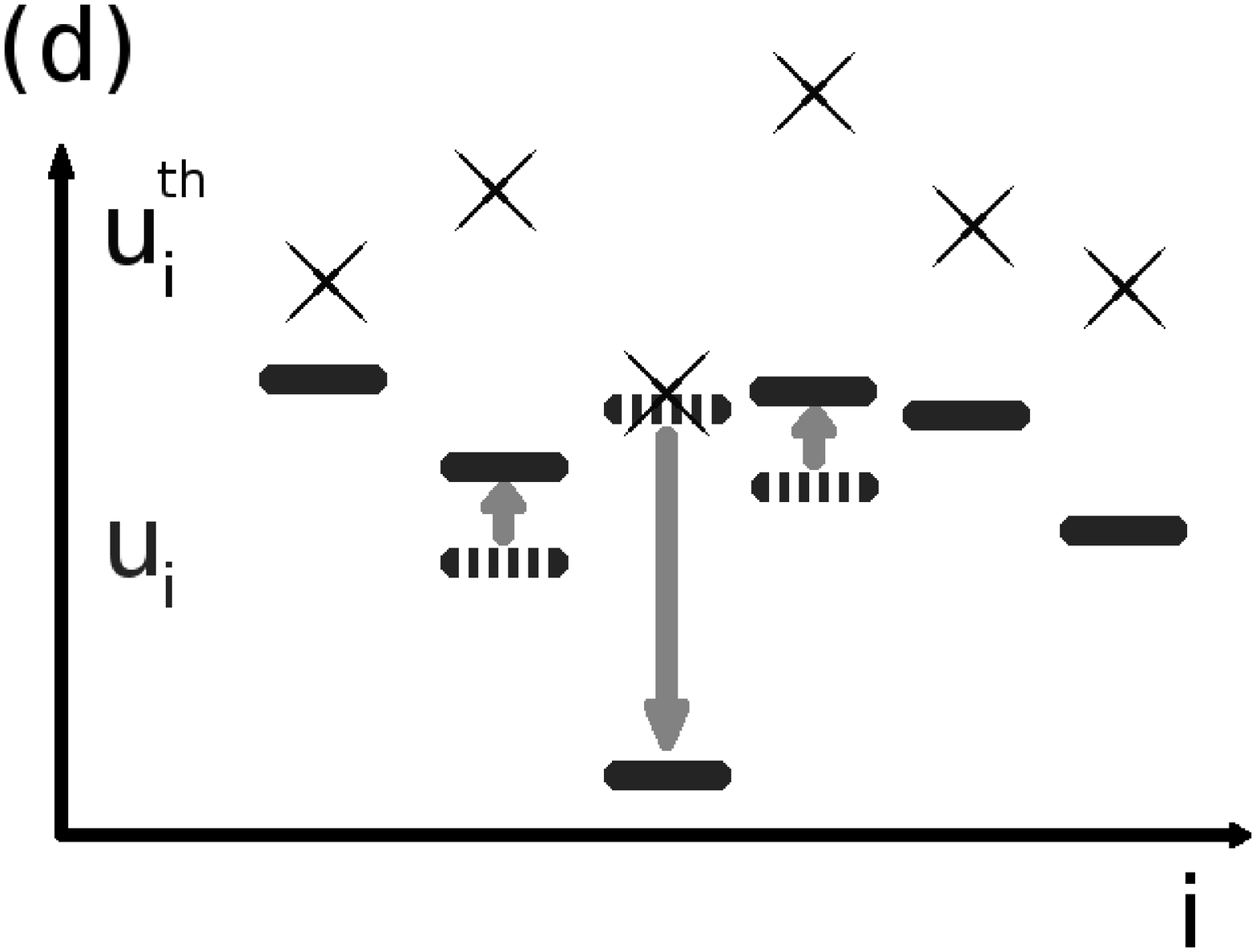}
	}
	
\caption{A sketch of the sliding situation that we are studying. (a) Two solid blocks slide against each other due to a constant driving 
between the top and bottom planes. Their relative velocity is $V$.
Dimensions of the blocks are $L_x$, $L_y$ in the sliding plane, and $L_z$ perpendicularly to it. (b) The solids in (a) are replaced by a rigid surface and an array of small blocks joined by springs with stiffness $k_0$. Driving acts on each block through a spring
of stiffness $k_1$. Note that the value of $k_1$ is inversely proportional to $L_z$ in (a). (c) The values of friction forces $u_i$ and maximum forces $u^{th}_i$ at each site are shown for the OFC* model. Friction forces increase uniformly (as indicated by the arrows) as driving proceeds. (d) The prescription as one $u_i$ reaches the local threshold: $u_i$ decreases by one unit, and the neighbors increase their $u$ values a quantity $\alpha$.}
\label{f1}
\end{figure}

Our starting point is the cellular automata model proposed by Olami, Feder, and Christensen\cite{ofc} (OFC) to describe seismicity.
The OFC model considers a set of real valued variables 
$u_i$ where $i$ indicates the position in a two dimensional lattice. 
$u_i$ is interpreted as the force that a rigid substrate exerts on a solid block at position $i$, and it represents the
local stress between the sliding plates (see Fig. \ref{f1}).
The system is driven by uniformly increasing the values of $u_i$ with time at a rate $V$, simulating the tectonic loading of the plates. Every time one of the variables $u_i$ reaches 
a maximum value (ordinarily set to an uniform, dimensionless value of 1), the local stress $u_i$ is `discharged' by setting it to zero. 
This local stress drop $\Delta u$ produces a stress increase onto neighbor blocks according to $u_j \to u_j + \alpha\Delta u$, where $j$ indicates a neighbor site to $i$. The value of $\alpha$
can vary between $0$ and $\alpha_c\equiv 1/Z$, $Z$ being the number of neighbors in the lattice. We will refer only to the case of a square lattice, so $Z=4$, $\alpha_c=1/4$. The case $\alpha=\alpha_c$ is called the conservative case, whereas $\alpha<\alpha_c$ are non-conservative cases. In the mechanical interpretation of the model, the value of $\alpha$ is related with the stiffness of the springs that inter-connect the blocks, and springs that push from the blocks at velocity $V$ (see Fig. \ref{f1}(b)). The actual relation is $\alpha=k_0/(k_1+4k_0)$.
A discharge event can produce the overpass of the maximum local stress on one, or more than one neighbor and in this way a large cascade can be generated. This cascade is called an event, and is identified with an individual earthquake (note that the complete cascade is assumed to occur at constant time, namely, earthquakes are instantaneous). The size $S$ of an event is defined as the sum of all discharges that compose the event, and the magnitude is defined 
as $M=\frac{2}{3}\log_{10} S$, so to match (up to an additive constant) the usual definition used in geophysics \cite{scholz}.

The OFC model is typically simulated using open boundary conditions, otherwise the spatial homogeneity
generated by the use of periodic conditions induces a strong global synchronization in the model.
The OFC model displays an exponential decay of number of events as a function of magnitude (or a power law decay if expressed in terms of $S$) compatible with a GR law of the form $N(M)\sim 10^{-bN}$. The $b$-value depends on the value of $\alpha$. Realistic values of $b$ are obtained for $\alpha \simeq 0.2$. 

\subsection{OFC* model}
The maximum values that the variables $u_i$ can withstand in the OFC model, are set to a constant, uniform value of 1. Having in mind a realistic situation
of a heterogeneous fault, with the constitutive materials having different properties at different positions, it becomes natural to consider a case in  which the threshold values are not constant but have some spatial variation. This is the first modification we introduce into the OFC model.
In concrete, the values of the local thresholds are called $u_i^{th}$,  and we draw them from some random distribution. Each time $u_i$ overpasses the local threshold $u_i^{th}$, $u_i$ is updated to a new value. We use the update rule $u_i \to u_i-1$, i.e, we implement a prescription of a unitary local stress drop. Upon this drop of the local stress, the values of $u$ on neighbor sites are updated as before, namely $u_j \to u_j + \alpha$, for $j$ neighbor to $i$ (see Fig. \ref{f1}(c-d)). 
Every time $u_i$ is updated, a new value is assigned to the local threshold $u_i^{th}$, taken from its original random distribution. This prescription is justified on the same physical arguments as before, since the sliding pieces can reasonably be thought as finding different maximum strengths as sliding proceeds. 
We refer to the OFC model with this modification as the OFC* model. 

The results obtained with the OFC* model differ qualitatively from those obtained with the OFC model.  In particular, the $b$ exponent becomes $\alpha$-independent, taking a value around $b\simeq 0.4$ (yet quite unrealistic for earthquakes). Interestingly, now the maximum avalanche size is controlled by the value of $\alpha$, diverging for $\alpha\to\alpha_c$ (assuming an infinite spatial extent of the system).  This may be considered a weak point, since it seems to indicate that is not sufficient to have an infinitely large system to have critical behavior, the condition $\alpha\to\alpha_c$ is also needed. 
However, we will see in Section V that this argument is flawed, as the large system size limit also implies $\L_z\to\infty$, and in this limit we naturally obtain that 
$\alpha\to \alpha_c$.

\subsection{OFCR model}
The second and crucial modification we made on the OFC model is the introduction of internal relaxation effects. We will not repeat here the arguments that justify its introduction, but let us just mention the following fact that points to the necessity of these effects: In the OFC, or OFC* models, a
sudden stop of the tectonic loading makes all seismic activity to cease at once. This is unrealistic, as for instance the processes that trigger aftershocks after a main event depend on the rearrangements that occur in the fault after the main shock, and are not directly related to tectonic loading. In other words, dynamical effects within the faults should be taken into account in order to have a realistic description of the seismic process. This is what we did by introducing the internal relaxation mechanism into the OFC* model. As this relaxation is controlled by a parameter $R$, we will call this the OFCR model. The concrete implementation prescribes that the evolution of the variables $u_i$ is given by

\begin{equation}
\frac{du_i}{dt}=R \left(\nabla^2 u\right)_i +V
\label{us}
\end{equation}   
namely, in addition to the external loading (represented by the $V$ term) there is a tendency to make the values of $u_i$ progressively more uniform in the system. This relaxation term produces the appearance of aftershocks and realistic friction properties in the system. In addition, the exponent of the GR law is modified, acquiring a value consistent with observations.\cite{jagla_kolton,jagla_ofc}

For the rest of the paper it will be important to consider a couple of variations of the previous relaxation mechanism. One of them is to consider relaxation in a ``mean field" manner. By this we understand that instead of the sum over the neighbor sites implied by the Laplacian operator in Eq. (\ref{us}) we take the average over the whole system, i.e, noting $\bar u$ the mean value of $u_i$, we have 

\begin{equation}
\frac{du_i}{dt}=-4R (u_i-\bar u) +V
\label{us2}
\end{equation}
(the factor of 4 is included to have a more direct comparison between the $R$ parameters in the two equations). 
The consideration of relaxation in mean field is important because it allows a very efficient determination of the next site that becomes unstable (i.e. the first site for which $u_i$ reaches $u_i^{th}$), as Eq. (\ref{us2}) can be solved analytically, whereas the same determination for Eq. (\ref{us}) requires a time consuming integration scheme. Results using both schemes of relaxation do not differ substantially. A comparison between the two schemes was presented in [5].

The second special case to be considered is the limit of ``infinite relaxation" defined by the condition $R/V\to\infty$. In this limit, there is a clear cut separation between events that are triggered by the tectonic loading $V$ term in Eqs. (\ref{us}) or (\ref{us2}) and those that are triggered by the relaxation term proportional to $R$. In fact, in this case events separate naturally into ``clusters". In between successive clusters, the spatial distribution of $u_i$ is uniform, and the $V$ term acts, moving up all $u_i$ uniformly until the lowest $u^{th}$ is reached. At this point, an event is triggered that leaves the $u$'s in a non-uniform state, over which the relaxation term acts. Under this action, aftershocks are triggered until the spatial distribution of $u$'s becomes uniform again, and the process is repeated.

\section{Background material on the $R=0$ case}

We present in this section a few reference results for the case of no relaxation, but with variable thresholds, namely, what we have called the OFC* model.

\subsection{Mapping between elastic interfaces in random media and the OFC* model}

First of all we show that the OFC* model can be mapped onto the problem of a two-dimensional elastic interface close to the depinning transition.  

Let us consider an ensemble of block positions $h_i$ identifying a two dimensional interface. The energy  of a given interface can be written as the sum of three contributions:

 (a) the elastic energy between nearest neighbors  blocks:  $ E_1=\frac{ k_0}{2 }   \sum_{\langle i, j\rangle} (h_i - h_j)^2 $,
 where  $h_i, h_j$ are the positions of the blocks $i$ and $j$, and  $k_0$ is the ``surface tension" of the elastic interface.

 (b) the energy of tectonic loading:  $ E_2=\sum_{i}  \frac{  k_1}{2}  (h_i - V t)^2$, 
 where $V$ measures the rate  and $k_1$ is the stiffness of the loading.
 
  (c) the random energy associated with the block position $h_i$, that is responsible for the pinning of the blocks.
  
 Elastic interactions and  tectonic  loading  generate a force $F_i=-\partial_{h_i} ( E_1+E_2)$  which competes with the pinning force $F_{\text{pinn}}(h_i)$ induced by the random energy.  The force $F_i$ acting on each block increases with time because of the loading. When $F_i$ overcomes $F_{\text{pinn}}(h_i)$ , the block $i$ becomes unstable and its position  is increased by one ($h_i \to h_i+1$). As a consequence  (i) the force $F_i$ drops to $F_i - (4 k_0 +k_1)$,  (ii) a new pinning  force is associated to the new position $h_i+1$ and  (iii) the force acting on the neighboring blocks  increases $F_j \to F_j +k_0$.   

 The dynamics of this interface in presence of a disordered landscape  can be easily identified with OFC* dynamics in presence of variable thresholds, the stress $u_i$ being $F_i/(k_1+4 k_0)$ and the control parameter $\alpha$ being $k_0/(k_1+4 k_0)$.

  \subsection{Review of the depinning transition}

The zero temperature motion of an elastic interface in random media has been studied in detail in the last two decades \cite{nattermann, fisher, wiese, kolton}. Two different  dynamical protocols are possible: (i) { \em constant velocity}: which corresponds to the OFC* model with a tectonic loading, or (ii) {\em constant force}: where  the parabolic tectonic loading, $E_2$, is replaced by the action of a constant and uniform stress $\sigma$, and the force acting on the block $i$ writes $F_i = (-\partial_{h_i}  E_1) + 4 k_0 \sigma$.

\subsubsection{Constant force}

For a given (and low) $\sigma$ the interface is pinned in a metastable state for which  the forces acting on all the blocks are smaller then the forces induced by the pinning centers.   When $\sigma$ is increased of an infinitesimal amount $\delta \sigma$ two things can happen: (i) either the metastable state remains metastable, or (ii) the metastable state becomes unstable and moves to a new metastable position. The distance between the old pinned position and the new one can be measured as the volume $S$ included between the two consecutive metastable states. This volume represents the size, $S$, of the event.   A critical threshold, $\sigma_c$ exists, above which there are no metastable states, and the interface evolves indefinitely. The transition between the pinned phase,  $\sigma< \sigma_c$, and the moving phase, $\sigma>\sigma_c$,  is called depinning.

Results of simulations at different values of $\sigma <\sigma_c$  show  that the distribution of events follow a power law,  with a large size cutoff,  $S_{\max}(\sigma)$ and an exponent $\tau>1$,
\begin{equation}
N(S)\propto \frac{1}{S^{\tau}}\,  f\left(\frac{S}{S_{\max}(\sigma)}\right). \label{scaling1}
\end{equation}
The exponent $\tau$ used by  the statistical mechanics community is related to the exponent $b$  typically used by geophysicists when looking at real earthquakes.  The relation between them is given by $\tau=1+2b/3$.
 We will refer mainly to values of $\tau$, from now on. The function $ f(S/S_{\max})$ has a fast decay when $S \gg S_{\max}$ and the characteristic large size, $S_{\max}(\sigma)$ diverges as $\sigma$ reaches a critical threshold $\sigma_c$.  The divergence of $S_{\max}$ goes with the divergence of the  linear size of the maximal avalanche $L_{\max}$. $L_{\max}$ is often seen as the ``growing correlation length" associated to this dynamical phase transition and,  close to $\sigma_c$,  $L_{\max}$ behaves as $   (\sigma_c-\sigma)^{-\nu} $ where, the positive exponent $\nu$ is defined in analogy with the correlation length exponent of equilibrium critical phenomena.

Large avalanches occur because the  system is ``organized" on large distances, another  consequence of this ``organization" is that the interface is a correlated object  displaying a power law roughness. A  positive roughness exponent $\zeta$  can be defined from  the growth of the distance between blocks  $|h_i-h_j| \sim |i-j|^\zeta$. An argument based on a symmetry of the system (statistical tilt symmetry)  \cite{fisher, wiese, kolton} allows to relate the exponent $\nu$ to the exponent $\zeta$, yielding
\begin{equation}
\nu=\frac{1}{2-\zeta}.
\label{tilt}
\end{equation}
From basic dimensional arguments we can  write
\begin{equation}
\label{smax}
S_{\max} \sim L_{\max}^{2+\zeta}\sim   (\sigma_c-\sigma)^{-\nu (2+\zeta)}  \sim  (\sigma_c-\sigma)^{-\frac{2+\zeta}{2-\zeta}}
\end{equation}

An important scaling relation between $\tau$ and $\zeta$ can be established if we compute the average size of an event $\bar{S}$. On one side, using the scaling form of Eq. \ref{scaling1},  this quantity can be written in terms of the cutoff size, in the form $\bar S \propto S_{\max}^{2-\tau}$,  as long as $1<\tau<2$ which is always the case here. On the other side we observe that the displacement of the center of mass $x_{\text{CM}}$
of the interface  when the stress in increased by an infinitesimal amount can be written as,  $x_{\text{CM}}(\sigma +\delta \sigma)-x_{\text{CM}}(\sigma)=\bar S N_\sigma(\delta \sigma)/(L_x L_y) $, where $N_\sigma(\delta \sigma)$ is the number of events when the stress jumps from $\sigma$ to $\sigma+\delta \sigma$.  If this number does not diverge when the threshold is approached (as confirmed by all simulations), then we can power expand, and to leading order we have  $N_\sigma(\delta \sigma)\sim c_1 \delta \sigma L_x L_y +\ldots$. Thus, it follows  that the average size of an event is proportional to $\bar S \sim \delta x_{\text{CM}}/\delta \sigma\equiv \chi$, the susceptibility of the interface. As the critical threshold is approached  the interface moves more and more,  the position of the interface grows as $x_{CM}(\sigma)\sim L_{\max}^\zeta$ and thus the susceptibility diverges as $ \chi  \sim (\sigma_c-\sigma)^{-\nu \zeta-1} $.  We thus obtain
\begin{equation}
\nu(2+\zeta)(2-\tau)= \nu \zeta +1 \;\;\;\; \Longrightarrow\;\;\;\; \tau =2- \frac{\zeta +\frac{1}{\nu}}{2+\zeta}.
\end{equation}
Using the statistical tilt symmetry relation (Eq. \ref{tilt}) we obtain
\begin{equation}
\tau=2-\frac{2}{2+\zeta}.
\label{tau}
\end{equation}

\subsubsection{Constant velocity}

At constant velocity, instead of a uniform stress, we have the parabolic potential $E_2$ centered at $V t$. The interface, embedded in this potential can be rough only at short length scales. The crossover between the short distance regime characterized by a rough interface and the long distance regime  with a flat interface can be determined by  dimensional analysis of the term $E_1 \sim L^{2 \zeta}$ against the term $E_2 \sim k_1 L^{2 \zeta+2 }$. The crossover length can be identified with the correlation length $L_{\text{max}}\sim 1/\sqrt{k_1}$. Using the statistical tilt symmetry  of elastic systems it is possible to show that this identification is actually correct. It is easy to verify that $\alpha_c-\alpha \sim k_1$, so that $\alpha_c-\alpha \simeq L_{\text{max}}^{-2} $.

\subsection{Results for the OFC* model}

\begin{figure}
	\includegraphics[width=0.5\textwidth]{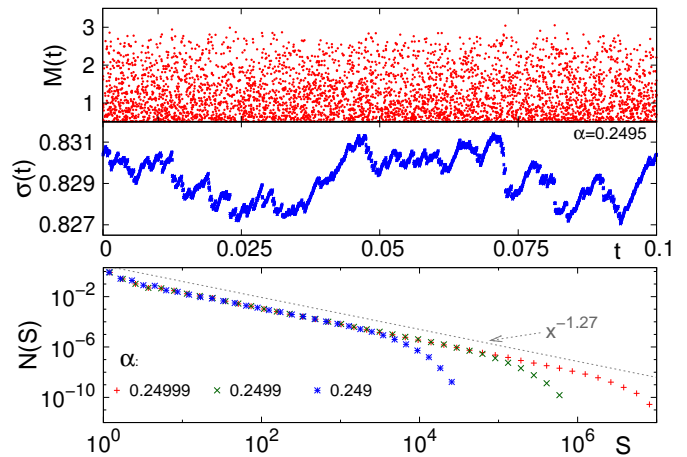}
\caption{Magnitude-time and stress-time plot for the OFC* model, and the event size distribution obtained for three different values of $\alpha$ as indicated. These
distributions are very well fitted by a power law with an exponential cut off.}
\label{f2}
\end{figure}

\begin{figure}
	\includegraphics[width=0.5\textwidth]{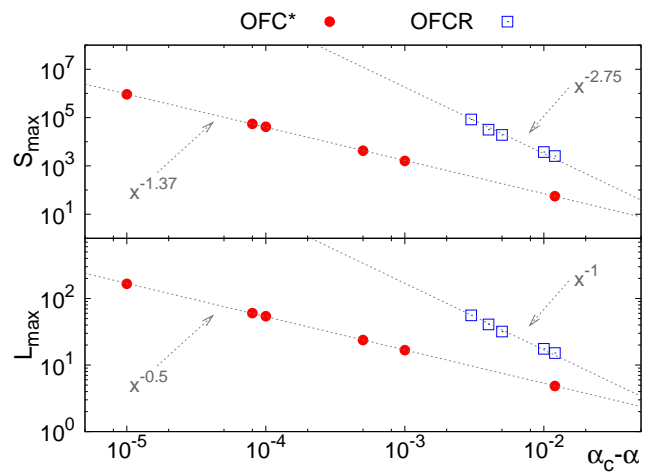}
\caption{Scaling of (a)$S_{\text{max}}$ and (b)$L_{\text{max}}$ with ($\alpha_c-\alpha$) for the OFC* and the OFCR models, shown in full and open symbols respectively.}
\label{f11}
\end{figure}

The previous arguments are consistent with the results of our simulations in the OFC* model. In Fig. \ref{f2} we show a sequence of individual events and the corresponding stress as a function of time for a fixed $\alpha$, and the event size distribution obtained for different values of $\alpha$. A value $\tau\simeq 1.27$ is systematically observed.
Also, from the distributions of $N(S)$ and the corresponding ones for $N(L)$ ($L$ is the typical linear size of the avalanches) we obtain the value of the large scale cut off $S_{\text{max}}$ and the maximum linear size $L_{{\max}}$ as a function of $\alpha$ for the OFC* model. Results are presented in Fig. \ref{f11}. 
We found $S_{\max} \sim (\alpha_c-\alpha)^{-\gamma}$  with $\gamma\simeq1.37$, which is consistent with the ratio $\gamma= 1+\zeta/2$ predicted by the 
theory of elastic interfaces of the previous section. We also find from Fig. \ref{f11} that $L_{\max}\sim (\alpha_c-\alpha)^{-1/2}$, showing that statistical tilt symmetry is satisfied by the OFC* model, as the mapping to the elastic interface problem could have anticipated.

\begin{figure}
\includegraphics[width=0.5\textwidth]{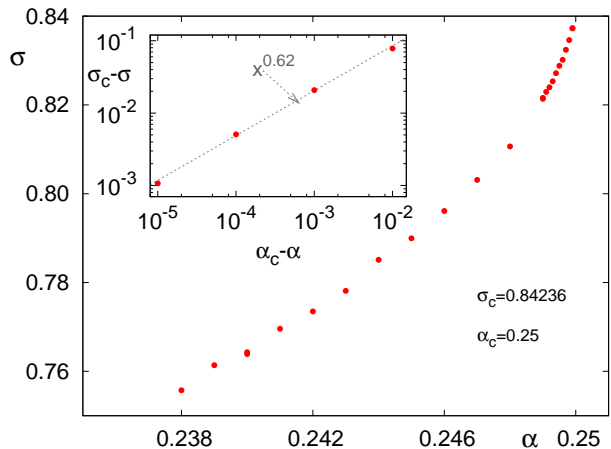}
\caption{Time averaged values of $\sigma$ as a function of the values of 
$\alpha$ for the OFC* model. The power law dependence close to $\alpha_c=0.25$, 
$\sigma_c\simeq 0.84236$ is emphasized in the logarithmic plot in the inset.}
\label{f3}
\end{figure}

For a fixed value of $\alpha$ we can follow the value of the average stress in the system (see Fig.  \ref{f2}), $\sigma\equiv \overline u_i$.  The stress settles around a mean value, with fluctuations typical of finite size effects. 
According to arguments in the previous section, the time averaged $\sigma$ vs. $\alpha$ is expected to have a critical behavior as $\alpha \to\alpha_c$ of the form $(\sigma_c-\sigma)\simeq (\alpha_c-\alpha)^{1/(2 \nu)}$ for some $\sigma_c$.  This is verified in our numerical simulations (Fig. \ref{f3}), as $1/(2 \nu) \simeq 0.62$.

The existence of a well defined average value of $\sigma$ for each value of $\alpha$ allows to recover the scaling relation
(Eq.(\ref{tau})) between $\tau$ and $\zeta$ using a slightly different argument. If an event of size $S$ occurs in the system, it produces a reduction of $\sigma$ by an amount proportional to $S\Delta\alpha/(L_xL_y)$ ($\Delta\alpha\equiv \alpha_c-\alpha$). If it is assumed (as it is in fact numerically verified) that the time interval between events does not become singular as $\Delta \alpha \to 0$, then there is a typical stress increment between events of the order of $1/(L_xL_y)$. To obtain a stationary mean value of $\sigma$, these two quantities have to compensate, on a temporal average, namely, we can write $\bar S\Delta\alpha\sim 1$ where with $\bar S$ we note the average value of the events. Putting  this estimation together with $\bar S \propto S_{\max}^{2-\tau}$ and using also Eq. (\ref{smax}), we re-obtain Eq. (\ref{tau}).

\begin{figure}
	\includegraphics[width=0.5\textwidth]{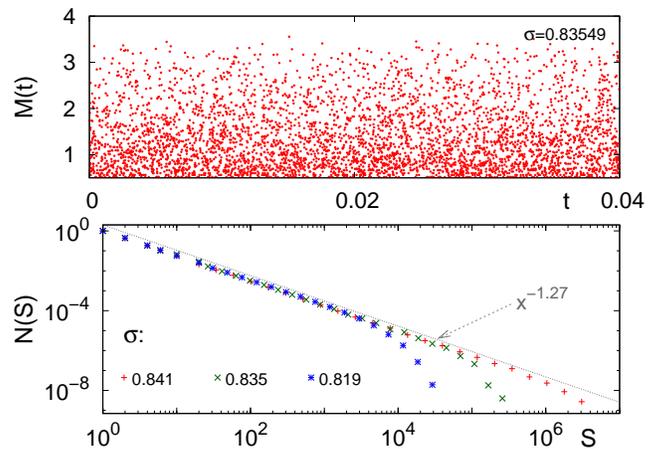}
\caption{Magnitude-time plot in the OFC* model, taking $V=0$, $\alpha=1/4$, and using the stress $\sigma$ as a control parameter, and event size distribution obtained for three different values of $\sigma$ as indicated. Note the similarity with Fig. \ref{f2}.}
\label{f4}
\end{figure}

Simulations of the OFC* model at $\alpha<\alpha_c$ and constant velocity are associated (in the mechanical analogy of the model) to driving by a finite stiffness spring at  a constant velocity.
The univocous relation between $\alpha$ and $\sigma$ allows also an alternative interpretation of the model, which is equivalent to driving the model at constant force.
Instead of applying a finite driving at fixed $\alpha<\alpha_c$, we set $\alpha=\alpha_c$, the driving velocity $V=0$, and fix the mean value of $\sigma$ from outside. 
A random site is chosen to initiate the avalanche.
After the avalanche is exhausted, the process is repeated. Note that $\sigma$ does not change in this process, and remains equal to the externally imposed value. In this way of doing the simulation, after a transient period, the avalanche size distribution becomes stationary in time, and its statistics coincides with that obtained fixing a non-critical $\alpha$ and having a finite $V$, as can be observed in Fig. \ref{f4}. The  expected behavior of $S_{\text{max}}$ as a function of $\sigma$ is given in Eq. (\ref{smax}). This behavior is consistent with our data for which $\nu (2+\zeta) \simeq 2.2$.

\section{OFCR model and the seismic cycle}

\begin{figure}
\includegraphics[width=0.5\textwidth]{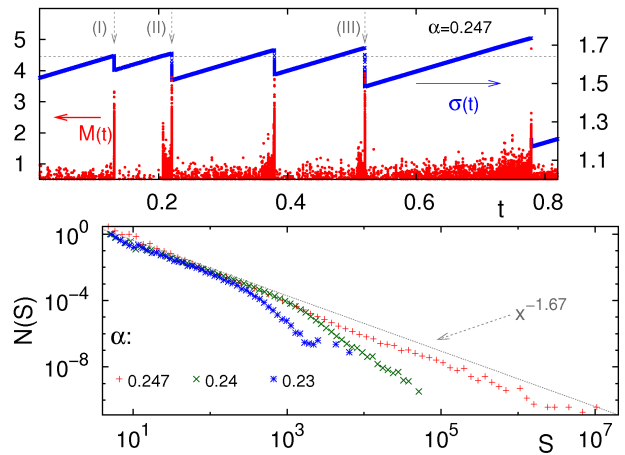}
\caption{(a)Magnitude-time and stress-time plot for the OFCR model. The probability of large events is strongly enhanced every time some threshold value of the stress (roughly indicated by the horizontal dotted line) is over passed. (b) The corresponding event size distribution for the whole sequence at different values of $\alpha$.}
\label{f5}
\end{figure}

\begin{figure}
\includegraphics[width=0.5\textwidth]{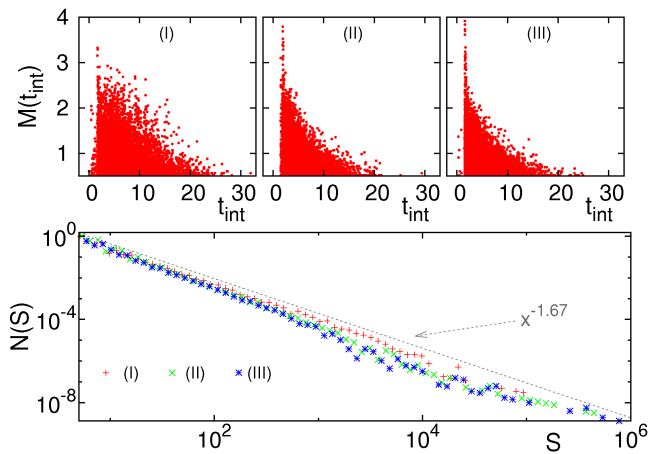}
\caption{(a)Magnitude-internal time ($t_{int}$) for some of the main clusters observed in Fig. \ref{f5}(a). $t_{int}$ is measured in units of the relaxation parameter, i.e, a whole cluster here occurs at a single horizontal coordinate in Fig. \ref{f5}(a). (b) the event size distribution of the individual clusters.}
\label{f6}
\end{figure}

In this section we begin to analyze results obtained in the presence of relaxation. We will work in the limit of infinite relaxation, and use the mean field implementation of relaxation. A typical sequence of events for this case is shown in Fig. \ref{f5}. We also indicate in that figure the evolution of $\sigma$ over the simulation. In the infinite relaxation case we will refer to an ``internal" and an ``external" temporal scale, which are totally decoupled. The external time scale is the scale of tectonic loading, and internal time scale is the scale of relaxation in Eq. (\ref{us}).
In Fig. \ref{f5} we plot events as a function of the external time scale. This means that at a fixed value of the horizontal coordinate, we find a whole ``cluster" of events. The events within some of these clusters, depicted as a function of the internal time scale are shown in Fig. \ref{f6}(a). 
One may consider a cluster as consistent of a main shock and all the aftershocks it produces. The clear cut identification of aftershocks as linked to some main event can be unambiguously made only in the present case of full separation of time scales.
Note however that within a cluster, the largest shock is not necessarily (nor even usually) the first one. In any case, we can consider a cluster as the full sequence of events that is committed to appear once the first event has been triggered, in the case in which tectonic loading is stopped after the initial event.
We emphasize that in the $R\to\infty$ case, the value of $\sigma$ shown in Fig. \ref{f5} is not only the average value of $u$ through the system, but it is also the actual value of every $u_i$.

In the results of Fig. \ref{f5} we see a systematic pattern of steady increase of $\sigma$, followed by abrupt drops. 
We interpret this cycle of smooth stress rising and sudden stress decrease as the manifestation in our model of a ``seismic cycle"\cite{scholz} . 
The concept of the seismic cycle suggests that stress on a fault smoothly builds up for a long period of time until is suddenly released in a large earthquake, and the process starts over. It leads to the idea that large earthquakes in a given region occur in a quasi-periodic manner. This idea, however, has been difficult to verify, on one side because of the large time intervals between large earthquakes, and also because deviations from periodicity could be so large as to transform a quasi-periodic sequence in something rather unpredictable. We notice however, that there have been a number of attempts (sometimes successful \cite{chile}) to infer the likelihood of a next large earthquake based on the record of previous large earthquakes in a particular region.

In the simulations, an alternation of large events and periods of rather low, slowly increasing activity is observed (see Fig. \ref{f5}). However, this
sequence is not periodic, instead it looks like a more or less random process.
In our case, we have the advantage to have access to all variables in the system, particularly to the instantaneous value of the stress, and thus we see that a large event occurs with large probability once some typical value of the stress (roughly indicated by the horizontal dotted line in Fig. \ref{f5}) has been over passed.
These observations point to the fact that there is some degree of predictability of the appearance of large earthquakes in the present model. However, we will not discuss here this issue, that we plan to elaborate in a forthcoming publication.

\begin{figure}
\includegraphics[width=0.5\textwidth]{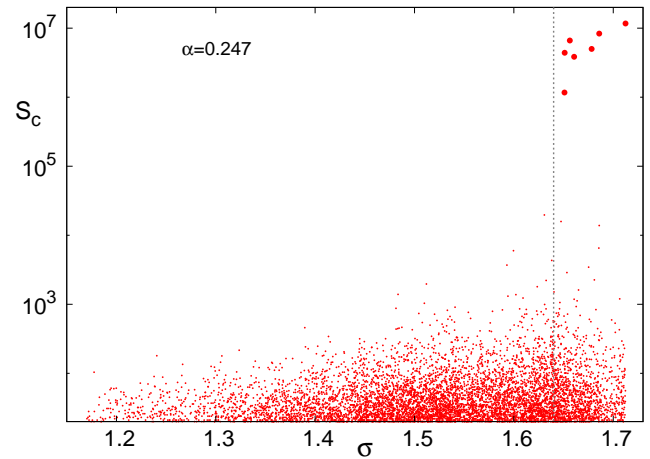}
\caption{The seismic moment of individual clusters $S_c$ vs. the initial value of stress at which they were triggered. The crossover stress (corresponding to the horizontal dotted line in Fig. \ref{f5}) is indicated by the vertical
dotted line. When stress becomes larger than this value there is some probability to trigger a very large cluster (those large clusters have been highlighted using a larger symbol size).}
\label{f7}
\end{figure}

The fluctuations of stress during the seismic cycle for the OFCR model are more fundamental than those observed for the OFC* model in the previous section. 
Insight into this issue is provided by taking the data in Fig. \ref{f5} and plotting the total seismic moment $S_c$ of each cluster (namely, the sum of the seismic moments of all event within a cluster) which is proportional to the stress drops observed in the curve of $\sigma(t)$ as a function of the value of stress at which the cluster is triggered. The result is shown in Fig. \ref{f7}.
We see that when $\sigma$ is small, the clusters are typically small also and the stress drop caused by these relatively small clusters is not able to compensate the stress increase caused by tectonic loading.  As $\sigma$ becomes larger than a certain value (indicated by the vertical dotted line in  Fig. \ref{f7}), there is a finite probability that a very large cluster is triggered. The size of these large clusters diverges as $\alpha\to\alpha_c$. These large events generate the abrupt decrease in stress observed in the previous Fig. \ref{f5}, re-initiating the seismic cycle.
This tells that the fluctuations of stress in the OFCR model are a necessary part of the dynamics of the model, which is eminently non-stationary, and the existence of a seismic cycle is the manifestation of that. 

\section{The macroscopic limit of the OFCR model}

There is a clear cut difference in the macroscopic limit of the OFC* and OFCR models that shows that the OFCR model is 
a candidate to be a realistic model for describing seismicity, whereas the OFC* model is not. This is related with the scaling of the
largest events observed for the two models in the macroscopic limit that we intend to discuss now.

This analysis starts from the key observation that the value of $\alpha_c-\alpha$ should be considered as proportional to $L_z^{-1}$, where
$L_z$ indicates the thickness of the slabs that slide against each other. This is clear by comparing Fig. \ref{f1}(a) and (b): the value of $k_1$ (and thus of $\alpha_c-\alpha=k_1/4(4k_1+k_0)$) is proportional to the restoring force the system exerts upon a given displacement of the top plane of the slab. For a fixed displacement, the restoring force is inversely proportional to $L_z$, namely $\alpha_c-\alpha\sim L_z^{-1}$.

In the macroscopic limit we should consider that $L$ ($L\equiv L_x=L_y$, the two lengths in the $x$-$y$ plane are taken equal for simplicity) and $L_z$ go to macroscopic values, although not necessarily with 
the same tendency. We will consider a case in which $L\gg L_z$, i.e., we maintain a sort of slab geometry. In this situation we want to calculate the  maximum spatial extent $L_{\max}$ of the earthquakes in the $x$-$y$ plane for OFC* and OFCR models. In the first case the statistical tilt symmetry (verified in the numerical simulations, see Fig. \ref{f11}(b)) implies that  $L_{\max}\sim 1/(\alpha_c-\alpha)^{1/2}$, from which $L_{\max}\sim L_z^{1/2}$. This means that as $L_z$ goes to infinity, the maximum spatial extent of the earthquakes becomes comparatively small. In other words, at large scales the sliding is smooth in the OFC* model, earthquakes do not survive the macroscopic limit. 

The situation is qualitatively different for the OFCR model. Here, numerical simulations indicate that the maximum spatial extent of the events has a different dependence with $(\alpha_c-\alpha)$. The results presented in Fig. \ref{f11}(b) indicate a dependence $L_{\max}\sim 1/(\alpha_c-\alpha)$, which in terms of $L_z$, can be written as $L_{\max}\sim L_z$.
This is a remarkable result. It indicates that no matter how large $L_z$ is, the maximum size of earthquakes is a sizable fraction of it. We can thus say that in this case, in the macroscopic limit earthquakes persist, and their maximum size scales as the system size itself. We think this is a key point 
to claim that this model can describe earthquakes in the macroscopic limit. 

The finding of sizable earthquakes with the size $L_z$ can be correlated with other properties of the model. It is accepted in the geophysics community, that systems in which earthquakes occur must have an effective friction law of the velocity weakening type\cite{scholz,vw}, namely, the average friction 
force must be a decreasing function of velocity, at least in a velocity range relevant for the process. Actually, we have shown elsewhere \cite{jagla_ofc} that the OFCR model does display velocity weakening, and in the origin of this behavior is again the existence of relaxation. 
It is immediate to show that such velocity weakening systems cannot sustain uniform sliding, and a stick-slip motion must necessarily occur for a sufficiently soft driving (i.e., for $\alpha\to\alpha_c$). 
Now if we assume a generic dependence of $L_{\max}$ on $L_z$ of the form $L_{\max}\sim L_z^\omega$, a value $\omega<1$ would correspond to an asymptotically smooth sliding, and since this cannot occur, we should have $\omega \ge 1$.
On the other hand, it
is very unlikely that $L_{\max}$ scales as a power of $L_z$ larger than one, because in a geometry of a slab with thickness $L_z$, a static perturbation at some point does not have an effect beyond some distance of the order of $L_z$. This takes us to the heuristic finding that the exponent effectively found in the numerical simulations, namely $\omega =1$
is the most natural result to be expected. 

A further question that we asked ourselves concerning this point is the following. The OFC* model displays a behavior with $\omega=1/2$, whereas our results for the OFCR model indicate $\omega=1$. How is the transition between these two exponents as the precise value of $R$ is varied between 0 (OFC*)
and a large value (OFCR)? Although this is an attractive question to explore, some investigation of the possibilities to provide an answer based on numerical simulations convinced us that we are not able to do so at present. We want to stress however our expectation that an answer to this question could give insight into the puzzling problem of creeping faults\cite{creeping}, namely a piece of a fault in which there is an almost complete absence of earthquakes and the sliding is smooth, while earthquakes occur in adjacent segments of the same fault. In our view this could be related to an abrupt transition between $\omega=1$ for a ``normal" segment, to a value $\omega<1$ in the creeping fault, and this abrupt transition could be driven by a smooth change in the amount of relaxation in the system. Clearly this issue is an open line of future research.

\section{Why $b\simeq 1$?}

One of the results that surprised us more concerning the OFCR model is the fact that it displays a realistic exponent $b\simeq 1$ ($\tau\simeq 1.67$), whereas the OFC* has $b\simeq 0.4$ ($\tau\simeq 1.27$). This is surprising because 
the inclusion of relaxation was aimed to generate aftershocks (what it does), and not to obtain a ``correct" GR law. So we believe there is a necessity to provide an explanation for this behavior. 

In section IIIB we have derived Eq. \ref{tau} linking the avalanche exponent $\tau$ with the roughness exponent $\zeta$ of the events of the OFC* model. The steps followed to arrive at Eq. \ref{tau} can be done also in the case of the OFCR model. The only important difference is that, as explained in the previous section, the scaling of $L_{\max} \sim (\alpha_c-\alpha)^{-\omega}$ has a different exponent for the OFCR case ($\omega=1$) compared to the OFC* case ($\omega=1/2$). The result that is obtained for a generic value of $\omega$ is
\begin{equation}
\tau=2-\frac{1/\omega}{2+\zeta}.
\label{tau2}
\end{equation}
This modified scaling relation suggests already a larger $\tau$ exponent in the OFCR case. In fact, assuming only a non-negative value for $\zeta$, we get $\tau>1.5$.  To determine  the value of $\zeta$ we need to characterize the internal structure of an avalanche. Four quantities can be used to this scope: (i) the already mentioned size $S$, which counts the total number of discharges, (ii) the area $A$ defined as the total number of blocks involved in the avalanche, (iii)  the ``duration" $T$ defined as the number of discharge steps that are necessary to exhaust the avalanche (note that sites that are critical at the same moment are discharged in parallel, in a single step), and  (iv) the linear size $L$ of the avalanche. Similarly to the size $S$, all these quantities are power law distributed up to a cut-off which depends on $\alpha_c-\alpha$. In the stationary regime, the study of the cut-off divergence, as the threshold $\alpha_c$ is approached, allows to determine the value of the critical exponents like $\zeta$, $\nu$, ... A different approach which does not rely on the stationary regime, consists in studying the behavior of $A$, $T$ and $L$ as a function of the size $S$. For large $S$ we expect a power law behavior with $\langle A(S) \rangle \sim S^{\kappa_A}$, $\langle T(S) \rangle \sim S^{\kappa_T}$ and $\langle L(S) \rangle \sim S^{\kappa_L}$, where $\langle A(S) \rangle$ is the area averaged over all avalanches of size $S$, $ \langle T(S) \rangle$ is the duration averaged over all avalanches of size $S$ and $ \langle L(S) \rangle$ is the linear size  averaged over all avalanches of size $S$. For the OFC* model,  scaling relations predict in two dimensions: $\kappa_A = \frac{2}{2+\zeta}\sim 0.73$,  $\kappa_T = \frac{z}{2+\zeta}\sim 0.57$ and  $\kappa_L = \frac{1}{2+\zeta}\sim 0.36$, where we have used  the value $z=1.56$ for the dynamical exponent.  Numerical simulations on the OFC* model agree with this predictions. More surprisingly also the avalanches of the OFCR model seem to have the same internal structure and  the same value of the $\kappa$ exponents. 
As a first consequence  if the value of $\zeta$  does not change appreciable from the value $\zeta\simeq 0.75$   in the absence of relaxation, through Eq. \ref{tau2}, we predict $\tau\simeq 1.64$, which is consistent with the value observed in the simulations (Fig. \ref{f5}). On the other hand the invariance of $\zeta$ when including relaxation gives a clue on the possible mechanism for the change of the $\tau$ exponent. In fact, if the new value of $\tau$ corresponds to the system being controlled by a different fixed point in the parameters space (i. e. a new universality class), we should expect that all exponents of the OFCR model are different from those of the OFC* model.  However, as explained by Durin and Zapperi\cite{creep_rupture} in the context of a magnetic problem, if the internal ($\kappa$) exponents of the avalanches do not change, it is likely that the new $\tau$ exponent originates as an effect of non-stationarity. 
If there is in the model a parameter that samples a wide distribution of values, the complete event size distribution can be generated as an integration of partial distributions restricted to particular values of the parameter. If, for instance, this parameter controls the position of the cut off of the distribution, such an integration produces a larger effective $\tau$ exponent, but does not affect the exponents $\kappa_A$, $\kappa_T$ and $\kappa_L$ characterizing the internal structure of the avalanches. 

A parameter that widely fluctuates in the OFCR model is the stress $\sigma$.
As $\sigma$ controls the position of the cutoff of the distribution, 
an integration over all $\sigma$ values could be the responsible of an overall power law with a larger decay exponent. If this explanation is correct, the event size distribution restricted to some small stress interval should display a $\tau\simeq 1.27$, with a cutoff that is stress-dependent. However this is not the case.
In Fig. \ref{f6}(b) we have shown event size distributions restricted to the events in some of the individual clusters observed in Fig. \ref{f5} (and thus characterized by a single value of $\sigma$ at triggering). We see for each individual cluster an event size distribution with the modified exponent $\tau\simeq 1.67$, indicating that $\sigma$ is not the parameter over which the effective integration takes place. 
Some other less obvious scheme of ``integration over a parameter" must be responsible for the change in exponent.

\begin{figure}
        \subfigure{
        \includegraphics[width=0.45\textwidth]{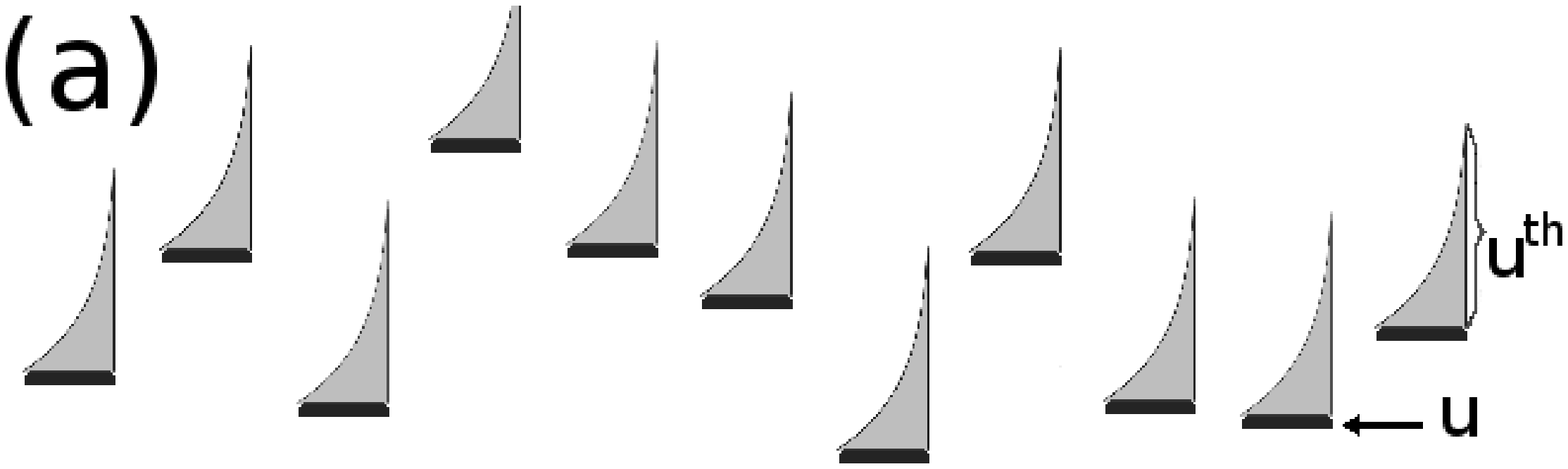}
        }
        \subfigure{
        \includegraphics[width=0.45\textwidth]{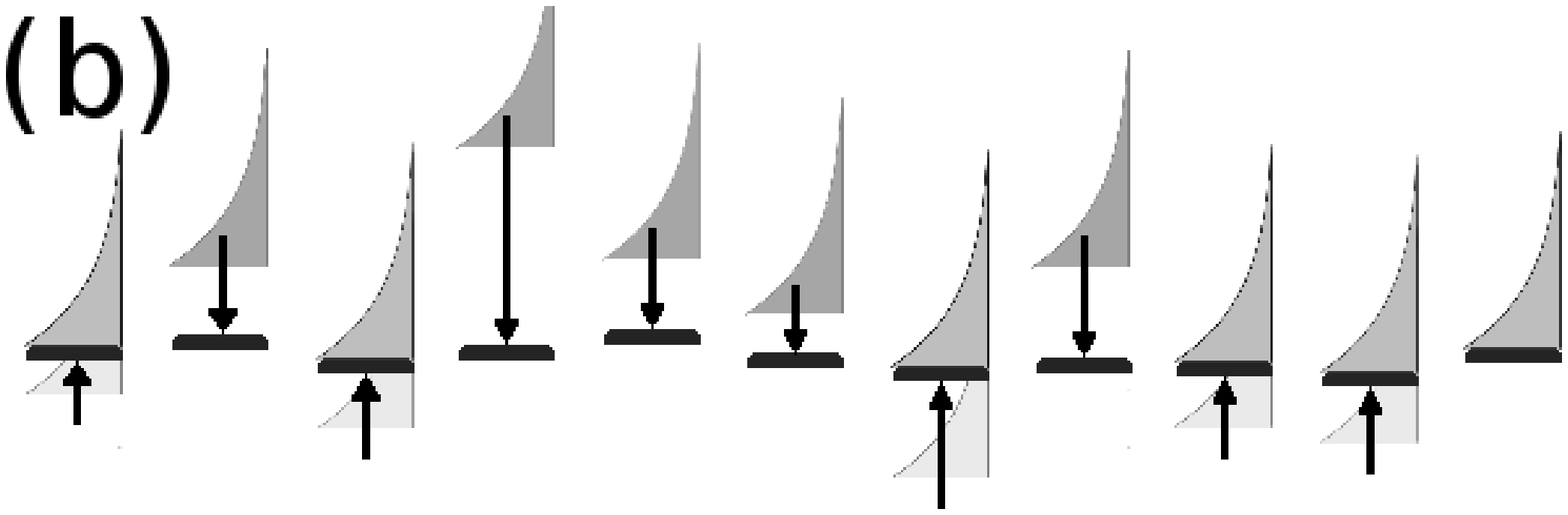}
        }

\caption{(a) A typical distribution of forces $u$ and thresholds $u^{th}$ in the absence of relaxation. For $u^{th}$ we show in gray the distribution from which each value is chosen, and not the values themselves. In the presence of relaxation, the configuration is modified as indicated in (b): sites for which relaxation produces an increase of $u_i$ can trigger aftershocks. If not, the distribution of $u^{th}$ continues to be exponential. Sites for which relaxation produces a decrease of $u_i$ get a gap between $u$ and the lowest possible $u^{th}$, and make these sites have a lower probability to be destabilized upon receiving load from neighbors.
}
\label{f9}
\end{figure}

We have not been able to identify clearly in the OFCR model the parameter that produces the effective integration. However, we have advanced along this line by doing the following simplified analysis.
Let us consider a distribution of values of $u$ and $u^{th}$. The thresholds are assumed to have an exponential distribution. 
The typical distribution of $u$ and $u^{th}$ in the absence of relaxation is sketched in Fig. \ref{f9}(a).  Note that instead of showing the actual values of $u^{th}$, we show the distribution from which they are chosen. 
In the presence of relaxation, the values of the forces move as indicated in Fig. \ref{f9}(b). 
In this process, any site for which $u_i<\sigma$ increases its value. As far as the actual value of $u^{th}$ is not reached, the probability distribution of $u^{th}$  continues to be exponential above $u_i$. If $u_i$ reaches $u_i^{th}$ an aftershock is triggered. On the other hand, sites for which $u_i >\sigma$ evolve decreasing its $u_i$ value because of relaxation.

\begin{figure}
\includegraphics[width=0.45\textwidth]{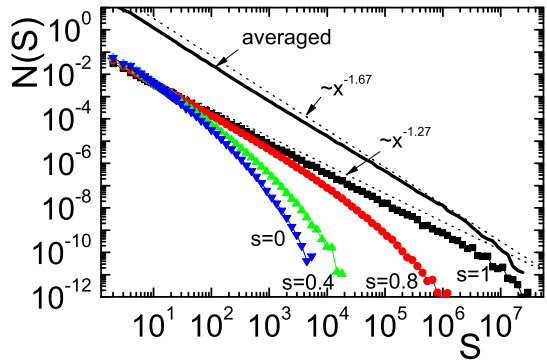}
\caption{Avalanche size distribution for an uncorrelated distribution of $u$'s and a shrink parameter $s$, as indicated.
The distribution of $u$'s is flat between $\bar u\pm \delta u$, with $\bar u=1.29$, and $\delta u=1$.
The result of an integration over the parameter $0<s<1$ is shown by the line labeled ``averaged" (vertically shifted, for convenience).
As a reference, power laws with exponents 1.27 and 1.67 are also plotted.
}
\label{u1}
\end{figure}

To understand the effect of relaxation, and having in mind the previous scheme,  we do the following simulation.
We take uncorrelated values of $u$, from some distribution of width $\delta u$ and mean $\bar u$, and take thresholds from an exponential distribution above the corresponding $u$, to mimic the configuration in Fig. \ref{f9}(a). Then, the values of $u$ are shrunk towards $\sigma$ by a factor $s$ ($0<s<1$) to simulate the effect of relaxation, namely $u\to \sigma +s(u-\sigma)$. If in this process $u$ becomes larger than the corresponding $u^{th}$, the threshold is chosen again. 
On this final configuration we destabilize a site at random and measure the size of the avalanche that is generated. Statistics is collected on many realizations of the process. Results are shown in Fig. \ref{u1}.
For $s=1$ (i.e., no relaxation) we get a power law distribution with an exponent close to (actually, a bit larger than) the normal value $\tau\simeq 1.27$. As $s$ is taken lower than one, the distribution becomes steeper. 
The qualitative reason of this lies in the gaps that appear between the $u$ values and the corresponding thresholds (Fig. \ref{f9}(b)): sites with larger gaps become less probable to be activated upon receiving load from a neighbor.

By choosing $s$ from a flat distribution between 0 and 1 (i.e., doing an integration over the value of $s$), a power law decay is recovered, with an exponent similar to that found with the full OFCR model. 
In a full simulation with the OFCR model, one can imagine a situation in which each event occurs at a particular value of $s$, and an effective integration over this parameter occurs, generating the modified exponent. 
A full justification of this scenario is not trivial, and we expect to provide it elsewhere.
In any case, our conclusion is that an effective integration over an internal  parameter that is related to relaxation is the responsible for the change of the $\tau$ exponent in the OFCR model.

\section{Summary and conclusions}

In this paper we have elaborated upon the properties of a seismicity model that is based on the one proposed by Olami, Feder, and Christensen, and that
incorporates relaxation effects as a fundamental ingredient. In the past, this model was shown  to generate realistic sequences of events, in particular displaying aftershocks following an Omori law. It also generates a Gutenberg-Richter event size distribution with realistic power law behavior, and reproduces frictional properties experimentally observed in geological materials. 

Here we have focused on the following issues. First of all, we have shown that there is a natural notion of a ``seismic cycle" in the model. This occurs because the stress in the system has temporal variations that reflect the overall state of the plates, and we have found that mayor earthquakes can be expected only when this stress is larger than some minimum value. Second, we have made an analysis of the scaling of the largest events in the system upon system size increase. With the appropriate interpretation of the parameter $\alpha$ of the model, it was suggested that in fact the size of the largest events scales as the system size (in particular, with the thickness of the plate we are trying to model). This indicates that the model is able to describe earthquakes in the macroscopic limit. Note that this is not true in the absence of relaxation, since in this case the largest events scale with system size with a lower-than-one power. 

Finally we have tried to understand why relaxation is able to tune the exponent of the GR law to a realistic value around $b=1$ (or $\tau\simeq 1.67$). We provided evidence that the reason of this modification is mainly the creation of ``gaps" between the $u$ values and the corresponding thresholds $u^{th}$, originated in the existence of relaxation. 
Despite this qualitative understanding of the effect of relaxation, we have not been able to explain why the exponent of the GR law in the presence of relaxation seems to adjust systematically around the value $b\simeq 1$ (or $\tau\simeq 1.67$). 
Taken as a whole, the results presented in this paper reinforce our view that the relaxation mechanism introduced on some seismicity models provides an important tool to study seismic phenomena.

\section{Acknowledgments}
This research was financially supported by Consejo Nacional de Investigaciones Cient\'{\i}ficas y T\'ecnicas (CONICET), Argentina. Partial support from
grants PIP/112-2009-0100051 (CONICET, Argentina) and  the France-Argentina MINCYT-ECOS A08E03 are also acknowledged. Finally A. R. acknowledges support by ANR grant 09-BLAN-0097-02.


\begin{thebibliography}{4}
\bibitem{ofc_aft}The OFC model does display aftershocks, but they are of a different nature than those observed
experimentally, see the discussion in [6].
\end{thebibliography}
\end{document}